\documentclass[showpacs,preprintnumbers,amsmath,amssymb,
twocolumn,
aps,prb]{revtex4}
\usepackage{graphicx}

\begin{document}

\title{Electrical initialization and manipulation of electron spins\\
in an L-shaped strained $n$-InGaAs channel}
\author{Y. K. Kato}
\author{R. C. Myers}
\author{A. C. Gossard}
\author{D. D. Awschalom}
\affiliation{Center for Spintronics and Quantum Computation,
University of California, Santa Barbara, CA 93106}
\date{\today}
\begin{abstract}
An L-shaped strained $n$-InGaAs channel is used to polarize and subsequently 
rotate electron spins solely by means of electric fields. Precession of 
electrically-excited spins in the absence of applied magnetic fields is 
directly observed by Kerr rotation microscopy. In addition, in-plane and 
out-of-plane components of the spin polarization in the channel are 
simultaneously imaged. 
\end{abstract}
\maketitle

Spin-orbit effects offer a unique direction for the development of 
spintronics 
\cite{Wolf:2001,Semiconductor:2002,Zutic:2004} by 
allowing spin-dependent interactions that require neither magnetic fields 
nor magnetic materials. A number of proposals take advantage of the 
spin-orbit interaction in semiconductors to construct spintronic devices 
from nonmagnetic materials, 
\cite{Voskoboynikov:2000,Kiselev:2003,Ting:2003,Hall:2003,Governale:2003} 
and recent experiments have demonstrated various manifestations of the 
spin-orbit interaction in semiconductor systems. It has been shown that an 
effective internal magnetic field $\textbf{B}_{\text{int}} $, which is 
proportional to the drift velocity of electrons, exists in strained $n$-type 
GaAs and InGaAs, enabling coherent manipulation of electron spins without 
magnetic fields. \cite{Kato:2004:1} Current-induced spin polarization 
(CISP) \cite{Levitov:1985,Edelstein:1990,Aronov:1989} has 
been detected in these systems \cite{Kato:2004:2} and in a two-dimensional 
hole gas at a GaAs/AlGaAs heterojunction, \cite{Silov:2004} demonstrating 
the feasibility of all-electrical spin initialization. The spin Hall effect 
\cite{Dyakonov:1971,Murakami:2003} is another phenomenon 
originating from the spin-orbit interaction and has been observed in 
$n$-type unstrained GaAs and strained InGaAs, \cite{Kato:2004:3} as well as in 
a two-dimensional hole gas. \cite{Wunderlich:2005} 

Here we show that it is possible to combine CISP and 
$\textbf{B}_{\text{int}} $ in strained semiconductors to polarize and 
subsequently rotate electron spins using electric fields alone, without an 
applied magnetic field. A Kerr microscope images the steady-state spin 
polarization in an L-shaped channel of strained InGaAs in which the spins 
orient in one arm and precess in the other arm.

\begin{figure}[b]\includegraphics{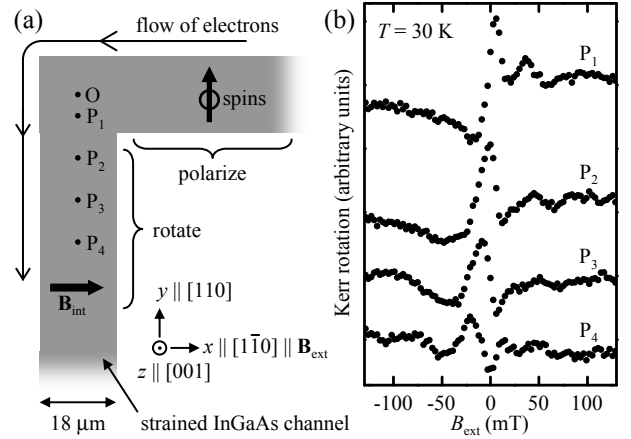}\caption{\label{fig1}
(a) Device schematic showing experiment geometry. Gray area is the InGaAs 
channel. The origin of the coordinate system used throughout is denoted by 
O, and the positions P$_{1}$ through P$_{4}$ are also indicated. The 
($x$,$y)$ coordinates in units of $\mu $m are as follows: P$_{1}\left( {0, - 5} 
\right)$, P$_{2}\left( {0, - 15} \right)$, P$_{3}\left( {0, - 25} 
\right)$, P$_{4}\left( {0, - 35} \right)$. (b) Voltage-induced Kerr 
rotation as a function of $B_{\text{ext}} $ for the four positions with $E = 
15$ mV~$\mu $m$^{ - 1}$. The curves are vertically offset for clarity.
}\end{figure}

Samples are fabricated from a wafer grown by molecular beam epitaxy on a 
semi-insulating (001) GaAs substrate (wafer ``E'' as described in previous 
works \cite{Kato:2004:1,Kato:2004:2,Kato:2004:3}), 
which consists of 500-nm thick $n$-In$_{0.07}$Ga$_{0.93}$As (Si-doped for $n = 
3\times 10^{16}$ cm$^{ - 3}$ to achieve long spin lifetimes 
\cite{Kikkawa:1998,Dzhioev:2002}) capped with 100 nm of undoped 
GaAs. The InGaAs layer shows anisotropic strain relaxation 
\cite{Kavanagh:1988} but residual strain is present as determined by 
X-ray diffraction at room temperature. Standard photolithography is employed 
to define an L-shaped InGaAs channel consisting of two arms, each with a 
width of 18 $\mu $m and a length of 200 $\mu $m, connecting at a right angle 
[Fig.~\ref{fig1}(a)]. The top 0.7 $\mu $m of the sample is chemically etched to form 
the mesa, and the two ends of the L-shaped channel are contacted by 
annealing the following metal layers, Au/Ni/Au/Ge/Ni, listed in order from 
the surface to the substrate. 

We consider the case where an electric field is applied such that the 
electrons flow from the horizontal arm to the vertical arm as shown in Fig.~\ref{fig1}(a). 
As the electrons flow to the $ - x$-direction in the horizontal arm, 
their spins are polarized in the $ + y$-direction because of CISP. After the 
electrons make the right angle turn and start flowing in the $ - 
y$-direction down the vertical arm, the electron spins will now see 
$\textbf{B}_{\text{int}} $ in the $ + x$-direction and start to precess. 
This will result in a non-zero component of spin polarization in the 
$z$-direction, allowing detection by Kerr rotation even at zero external 
magnetic fields. Previous measurements 
\cite{Kato:2004:1,Kato:2004:2} on a sample from the same 
wafer have shown anisotropy of CISP and $\textbf{B}_{\text{int}} $ for 
electric field \textbf{E} applied along the two crystal axes $\left[ {110} 
\right]$ and $\left[ {1\bar {1}0} \right]$. Larger CISP has been observed 
when $\textbf{E} \parallel \left[ {1\bar {1}0} \right]$, while 
$\textbf{B}_{\text{int}} $ is larger for $\textbf{E} \parallel \left[ {110} 
\right]$. In order to use this anisotropy to our advantage, we use the 
$[1\bar {1}0]$ axis for the horizontal arm along the $x$-axis and the $\left[ 
{110} \right]$ axis for the vertical arm along the $y$-axis.

Steady-state spin polarization is probed by a low-temperature scanning Kerr 
microscope in the Voigt geometry. 
\cite{Kato:2004:3,Stephens:2003} The sample is placed in a 
helium flow cryostat, and measurements are performed at a temperature $T = 
30$ K. A mode-locked Ti:sapphire laser is tuned to a wavelength of 873 nm, 
and a linearly polarized beam with an average power of 130 $\mu $W is 
directed along the $z$-axis (sample growth direction) and focused through an 
objective lens with a numerical aperture of 0.73 and a working distance of 
4.7 mm. The polarization axis of the reflected beam rotates by an amount 
proportional to the component of net spin polarization along the optical 
axis (Kerr rotation). A square wave voltage with a frequency $f_{E}$ = 1.169 
kHz is applied to the two contacts and establishes an alternating electric 
field with an average amplitude $E$. Kerr rotation is detected by a balanced 
photodiode bridge using a standard lock-in technique. We note that this 
measurement scheme also detects signal arising from electron flow opposite 
to that drawn in Fig.~\ref{fig1}(a), since the change in the Kerr rotation between 
positive and negative electric fields is measured. 

In Fig.~\ref{fig1}(b), Kerr rotation measurements for scans of the external magnetic 
field $B_{\text{ext}} $ applied parallel to the $x$-axis with $E = 15$ mV~$\mu 
$m$^{ - 1}$ is shown for various positions on the sample. Although the 
curves show complex behavior, qualitative explanation is possible with a 
simple model. We start by considering the steady-state spin polarization 
prepared in the horizontal channel. As was done in previous work, 
\cite{Kato:2004:2} CISP is modeled as a continuous orientation of spins 
along the $y$-axis. The $z$-component of the steady-state spin polarization vector 
$\textbf{S}\left( {x,y} \right)$ just prior to the right-angle turn is
\begin{equation}
S_z \left( {0,0} \right) = \int_0^\infty {dt\left[ {\gamma e^{ - t / \tau 
}\sin \left( {\omega _L t} \right)} \right]} = \frac{S_\text{0} \omega _L 
\tau }{\left( {\omega _L \tau } \right)^2 + 1},\label{eq1}
\end{equation}
where $\gamma $ is the rate of spin orientation onto the $y$-axis, $\tau $ is 
the inhomogeneous transverse spin lifetime, $\omega _L = {g\mu _B 
B_{\text{ext}} } \mathord{\left/ {\vphantom {{g\mu _B B_{\text{ext}} } \hbar 
}} \right. \kern-\nulldelimiterspace} \hbar $ is the electron Larmor 
frequency, $S_\text{0} \equiv \gamma \tau $ is the steady-state spin 
polarization achieved at $B_{\text{ext}} = 0$, $g$ is the effective electron 
g-factor, $\mu _{B}$ is the Bohr magneton, and $\hbar $ is the Planck 
constant. Similarly, the $y$-component of the polarization is
\begin{equation}
S_y \left( {0,0} \right) = \int_0^\infty {dt\left[ {\gamma e^{ - t / \tau 
}\cos \left( {\omega _L t} \right)} \right]} = \frac{S_\text{0} }{\left( 
{\omega _L \tau } \right)^2 + 1},\label{eq2}
\end{equation}
which is equivalent to the Hanle curve. \cite{Optical:1984}

Next, we consider the effect of $\textbf{B}_{\text{int}} $. If we pick a 
particular position on the $y$-axis, all spins will display the same precession 
phase regardless of their drift time since $\textbf{B}_{\text{int}} $ is 
proportional to the drift velocity of the electrons. \cite{Kato:2004:1} 
Then the $z$-component of spin polarization is
\begin{eqnarray}
\nonumber
 S_z \left( {0,y} \right) &=& S_z \left( {0,0} \right)\cos \left[ {\phi \left( 
{0,y} \right)} \right] + S_y \left( {0,0} \right)\sin \left[ {\phi \left( 
{0,y} \right)} \right] \\ 
 &=& S_\text{0} \frac{\omega _L \tau \cos \left[ {\phi \left( {0,y} \right)} 
\right] + \sin \left[ {\phi \left( {0,y} \right)} \right]}{\left( {\omega _L 
\tau } \right)^2 + 1}, \label{eq3}
\end{eqnarray}
where $\phi \left( {x,y} \right)$ is the phase that is acquired at the 
position $\left( {x,y} \right)$. We expect $\phi \left( {0,y} \right) = - 
\beta y / \hbar $, where $\beta $ is the spin-splitting coefficient defined 
in the previous work, \cite{Kato:2004:1} and this relation gives a 
spatial period of $2\pi \hbar / \beta = 37$~$\mu $m for this sample. As 
$\phi $ increases with more negative $y$, the magnetic field dependence 
deviates from the antisymmetric curve [Fig.~\ref{fig1}(b), top curve] to the more 
symmetric curve [Fig.~\ref{fig1}(b), second curve from top]. As the distance from the 
corner increases, such a model becomes too simplistic. First, since spins do 
not orient along the $ + y$-direction after the right-angle turn, the 
integration limits in Eq.~(\ref{eq1}) and (\ref{eq2}) become less accurate approximations. 
Second, the spin diffusion transverse to the electric field and the exact 
trajectory at the corner will affect the line shape. We also note that there 
are further complications due to the signal from the electron flow in the 
opposite direction, as mentioned previously. Although CISP in the vertical 
channel has smaller efficiency \cite{Kato:2004:2} and would not precess 
out of plane since it is parallel to $\textbf{B}_{\text{ext}} $, such a 
contribution can appear in the vicinity of the corner.

\begin{figure}\includegraphics{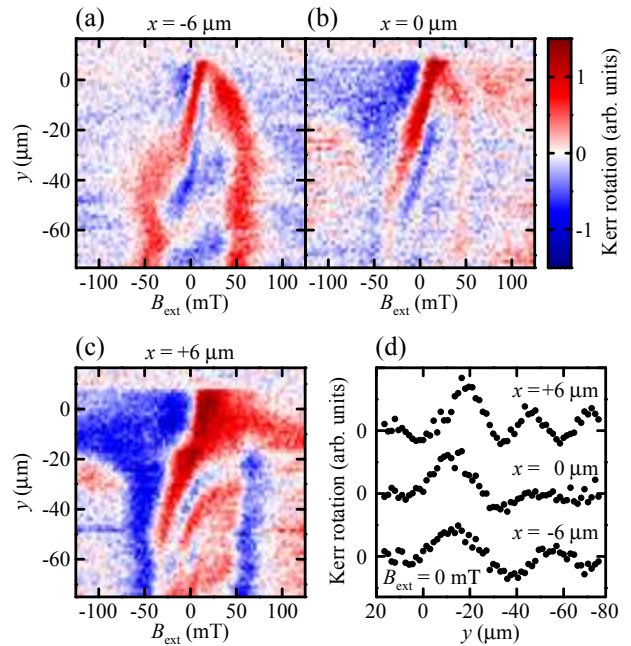}\caption{\label{fig2}
Kerr rotation as a function of $y$ and $B_{\text{ext}} $ with $E = 15$ mV~$\mu 
$m$^{ - 1}$ at $T = 30$ K for three line scans parallel to the $y$-axis at (a) 
$x = - 6$~$\mu $m, (b) $x = 0$~$\mu $m, and (c) $x = + 6$~$\mu $m. In order 
to improve the visibility of the features, we have subtracted the average 
value from each scan of $B_{\text{ext}} $. (d) Line cuts along the $y$-axis at 
$B_{\text{ext}} = 0$ mT from (a), (b), and (c). 
}\end{figure}
In Fig.~\ref{fig2}(a), (b), and (c), Kerr rotation is plotted as a function of $y$ and 
$B_{\text{ext}} $, for $x = - 6$~$\mu $m, $x = 0$~$\mu $m, and $x = + 6$~$\mu 
$m, respectively, with $E = 15$ mV~$\mu $m$^{ - 1}$. Qualitatively, the 
effect of $\textbf{B}_{\text{int}} $ can be identified in Fig.~\ref{fig2}(b), where 
the main positive peak at $B_{\text{ext}} > 0$ mT seen at $y = 0$~$\mu $m 
shifts to $B_{\text{ext}} < 0$ mT as $y$ becomes more negative. The electrons 
that have drifted towards $\left[ {\bar {1}\bar {1}0} \right]$ have been 
affected by $\textbf{B}_{\text{int}} $ so that $\textbf{B}_{\text{ext}} $ 
needs to be applied opposite to $\textbf{B}_{\text{int}} $ to cancel its 
effect. Line cuts at $B_{\text{ext}} = 0$ mT are plotted in Fig.~\ref{fig2}(d), 
showing oscillations as a function of position as expected from the simple 
sine function obtained by letting $\omega _L = 0$ in Eq.~(\ref{eq3}). This result 
provides unambiguous evidence that $\textbf{B}_{\text{ext}} $ is not 
required for CISP. In the previous work, $\textbf{B}_{\text{ext}} $ was used 
to rotate the polarization in order to allow detection by Kerr rotation, 
\cite{Kato:2004:2} while the data in Fig.~\ref{fig2}(d) takes advantage of 
$\textbf{B}_{\text{int}} $ instead. The oscillation period ranges from $\sim 
$30 to 50 $\mu $m for the three line cuts. While these values are consistent 
with the model, the $x$-dependence is currently not understood. In addition to 
CISP and subsequent precession caused by $\textbf{B}_{\text{int}} $, we also 
see the signatures of the spin Hall effect \cite{Kato:2004:3} in Fig.~\ref{fig2}(a) 
and (c). For $y < - 30$~$\mu $m and $B_{\text{ext}} $ around $\pm $50 
mT, positive peaks are seen in Fig.~\ref{fig2}(a) and negative peaks are seen in 
Fig.~\ref{fig2}(c). These peaks can also be seen for $x = 0$~$\mu $m [Fig.~\ref{fig2}(b)], 
although they are much weaker. Such signal contribution from the spin Hall 
effect can explain some features in individual $B_{\text{ext}} $ scans. For 
example, the small peak at $B_{\text{ext}} = 35$ mT in the top curve in Fig.~\ref{fig1}(b) 
is interpreted as arising from the spin Hall effect when the electrons 
flow from the vertical channel to the horizontal channel, since this peak 
smoothly shifts into one of the spin Hall peaks for more negative $y$ as seen 
in Fig.~\ref{fig2}(b). The spin Hall peak occurring at a smaller $B_{\text{ext}} $ 
can be explained by the reduced $y$-component of electron drift velocity in the 
vicinity of the corner.
\begin{figure}\includegraphics{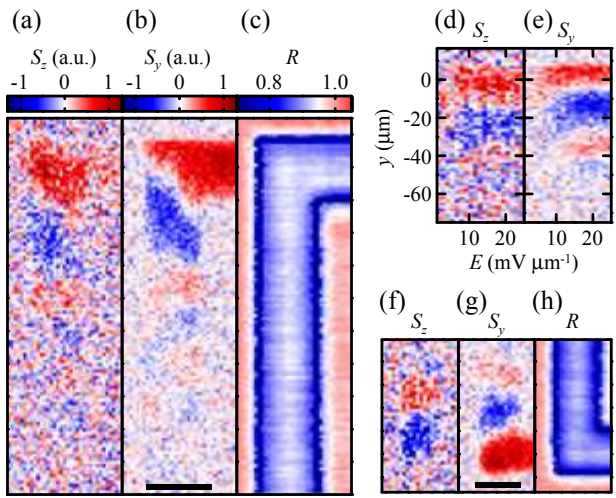}\caption{\label{fig3}
Images of (a) $S_{z}$, (b) $S_{y}$, and (c) $R$ for an L-shaped channel with $E = 
15$ mV~$\mu $m$^{ - 1}$. (d) and (e) are $S_{z}$ and $S_{y}$, respectively, as 
a function of $y$ and $E$ for a line scan parallel to the $y$-axis at $x = 0$~$\mu 
$m. Images of (f) $S_{z}$, (g) $S_{y}$, and (h) $R$ for an L-shaped channel which 
bends towards $\left[ {110} \right]$ with $E = 15$ mV~$\mu $m$^{ - 1}$. The 
scale bars are 20 $\mu $m and data are taken at $T = 30$ K. (a), (d), and 
(f) share the color scale for $S_{z}$; (b), (e), and (g) share the color 
scale for $S_{y}$; (c) and (h) share the color scale for $R$.
}\end{figure}

The spatial profile of the spin polarization in the L-shaped channel for 
$B_{\text{ext}} \sim 0$ mT can be mapped out with a technique previously 
used to image the spin Hall effect \cite{Kato:2004:3}. Here, 
$B_{\text{ext}} $ is sinusoidally modulated at $f_B = 3.3$ Hz with amplitude 
of 30 mT, and the signal is detected at $f_E \pm f_B $ and $f_E \pm 2f_B $. 
The former will be proportional to the first derivative of Eq.~(\ref{eq3}) with 
respect to $\omega _{L}$ and corresponds to $S_{y}$ at $B_{\text{ext}} = 
0$, while the latter is proportional to the second derivative which 
corresponds to $S_{z}$. We show $S_{z}$ and $S_{y}$ for $E = 15$ mV~$\mu $m$^{ 
- 1}$ in Fig.~\ref{fig3}(a) and (b), respectively, and the reflectivity $R$ is plotted 
in Fig.~\ref{fig3}(c). The oscillations are clearly resolved, with $S_{z}$ and 
$S_{y}$ being out-of-phase by $\pi $/2 as expected. The difference in 
oscillation period for different $x$ is reproduced in these images. The 
contributions from the spin Hall effect do not appear in these images since 
$\textbf{B}_{\text{int}} $ acts to depolarize spin polarization due to the 
spin Hall effect when $B_{\text{ext}} = 0$ mT. \cite{Kato:2004:3}

The electric field dependence of the oscillation is investigated by 
measuring line scans along the $y$-axis at $x = 0$~$\mu $m. In Fig.~\ref{fig3}(d) and 
(e), $S_{z}$ and $S_{y}$ are plotted, respectively, as a function of $y$ and $E$. 
The average electron velocity increases with increasing $E$, but the period of 
oscillations should not change because of the linear relation between 
$\textbf{B}_{\text{int}} $ and drift velocity. There is no drastic change 
with $E$ as expected, but for small values of $E$, the oscillation period slightly 
increases. This may be due to diffusion of spins transverse to $E$ playing a 
more important role for small values of $E$, as the above argument only holds 
for one-dimensional transport of spins.

Finally, we have also fabricated another sample on the same chip, with the 
vertical channel pointing to $\left[ {110} \right]$ ($ + y$-direction) 
instead of $\left[ {\bar {1}\bar {1}0} \right]$ ($ - y$-direction). The 
images for this sample are shown in Fig.~\ref{fig3}(f), (g) and (h), and show similar 
behavior for $S_{y}$ while the sign of the signal is inverted for $S_{z}$. 
This result is expected as $\textbf{B}_{\text{int}} $ changes sign when the 
electrons flow direction is changed from $\left[ {\bar {1}\bar {1}0} 
\right]$ to $\left[ {110} \right]$, and this demonstrates the flexibility to 
control $\textbf{B}_{\text{int}} $ through the shape of the channel. By 
combining channels with appropriate topology to utilize 
$\textbf{B}_{\text{int}} $ from different crystal axes, more complicated 
manipulation of spin states should be possible. 

We thank S. A. Wolf for helpful suggestions and acknowledge support from DARPA, 
DMEA, NSF, and ONR.

\end{document}